\documentclass[12pt]{article}
\usepackage{amssymb}
\newcommand{\be}{\begin{equation}}
\newcommand{\ee}{\end{equation}}
\newcommand{\bn}{\begin{eqnarray}}
\newcommand{\en}{\end{eqnarray}}

\newcommand{\dslash}{\partial\!\!\!/}
\newcommand{\aaslash}{A\!\!\!/}
\newcommand{\nn}{\nonumber}

\newcommand{\no}{\noindent}

\def\bea{\begin{eqnarray}}
\def\eea{\end{eqnarray}}
\begin{document}

\title{\textbf{Generalizing the Soldering procedure}}
\author{D. Dalmazi, A. de Souza Dutra and E. M. C. Abreu \\
%EndAName
\textit{UNESP - Campus de Guaratinguet\'a - DFQ } \\
\textit{Av. Dr. Ariberto Pereira da Cunha, 333} \\
\textit{CEP 12516-410 - Guaratinguet\'a - SP - Brazil.} \\
\textsf{E-mail: dalmazi@feg.unesp.br, dutra@feg.unesp.br }}
\maketitle

\begin{abstract}
We start this work by revisiting the problem of the soldering of two chiral
Schwinger models of opposite chiralities. We verify that, in contrast with
what one can conclude from the soldering literature, the usual sum of these
models is, in fact, gauge invariant and corresponds to a composite model,
where the component models are the vector and axial Schwinger models. As a
consequence, we reinterpret this formalism as a kind of degree of freedom
reduction mechanism. This result has led us to discover a second soldering
possibility giving rise to the axial Schwinger model. This new result is
seemingly rather general. We explore it here in the soldering of two
Maxwell-Chern-Simons theories with different masses.
\end{abstract}

\section{Introduction}

The study of chiral bosonic fields has arisen and proliferate mainly due to
its importance in the quantization of strings \cite{gross} and other
theoretical models \cite{grs}. This research in chiral bosonization has
begun many years back with the seminal paper of W. Siegel \cite{siegel}.
Floreanini-Jackiw have offered later some different solutions to the problem
of a single self-dual field \cite{fj} proposing a non-anomalous model. The
study of chiral bosons also plays an important role in the studies of the
quantum Hall effect \cite{ws}. The introduction of a soliton field as a
charge-creating field obeying one additional equation of motion leads to a
bosonization rule \cite{ggkrs}. The author of \cite{ms} has shown that the
method of coadjoint orbit \cite{orbitas}, when applied to a representation
of a group associated with a single affine Kac-Moody algebra, generates an
action for the chiral WZW model \cite{wzw}, which is a non-Abelian
generalization of the Floreanini and Jackiw (FJ) model.

The concept of soldering \cite{ms,adw,abw} has proved extremely
useful in different contexts. The soldering formalism essentially
combines two distinct Lagrangians manifesting dual aspects of some
symmetry to yield a new Lagrangian which is destituted of, or
rather hides, that symmetry. The quantum interference effects,
whether constructive or destructive, among the dual aspects of
symmetry, are thereby captured through this mechanism \cite {abw}.
The formalism introduced by M. Stone \cite{ms} could actually be
interpreted as a new method of dynamical mass generation
\cite{abw}. This technique parallels a similar phenomenon in two
dimensional field theory known as Schwinger mechanism \cite{schw}
that results from the interference between right and left massless
self-dual modes of chiral Schwinger model \cite{jr} of opposite
chiralities \cite{abw}. The result of the chiral interference
shows the presence of a massive vectorial mode, for the special
case where the Jackiw-Rajaraman regularization parameter is $a=1$
\cite{jr}, which is the value where the chiral theories have only
one massless excitation in the spectrum. This clearly shows that
the massive vector mode results from the interference between two
massless modes.

It was shown lately \cite{aw}, that in the soldering process of
two Siegel's \cite{siegel} modes (lefton and righton) coupled to a
gauge field \cite{gs}, this gauge field has decoupled from the
physical field. The final action describes a nonmover field (a
noton) at the classical level. The noton acquires dynamics upon
quantization. This field was introduced by Hull \cite {hull} to
cancel out the Siegel anomaly. It carries a representation of the
full diffeomorphism group, while its chiral components carry the
representation of the chiral diffeomorphism.

In the $3D$ case, the soldering mechanism was used to show the
result of fusing together two topologically massive modes
generated by the bosonization of two massive Thirring models with
opposite mass signatures in the long wave-length limit. The
bosonized modes, which are described by self and antiself dual
Chern-Simons models \cite{tpn,dj}, were then soldered into the two
massive modes of the $3D$ Proca model \cite{bw527}. In the $4D$
case, the soldering mechanism produced an explicitly dual and
covariant action as the result of the interference between two
Schwarz-Sen \cite{ss} actions displaying opposite aspects of the
electromagnetic duality \cite{bw527}.

In this work we revisit the problem of the soldering of two chiral Schwinger
models of opposite chiralities. Verifying that the usual sum of these models
is, in fact, gauge invariant and corresponds to a composite model, where the
component models are the vector and axial Schwinger models \cite{dutra}. So,
in this particular case, we show that it is not really necessary to use the
soldering mechanism to accomplish the gauge symmetry as supposed. As a
consequence, we reinterpret it as a kind of degree of freedom reduction
mechanism. This idea is then used in order to define other possible ways of
performing this soldering/fusion procedure.

\section{A Brief description of the soldering formalism}

The soldering formalism gives an useful bosonization scheme for Weyl
fermions, since a level one representation of LU(N) has an interpretation as
the Hilbert space for a free chiral fermion \cite{ps}. However, only Weyl
fermions can be analyzed in this way, since a $2D$ conformally invariant QFT
has separated right and left current algebras. In other words, it is trivial
to make a (free) Dirac fermion from two (free) Weyl fermions with opposite
chiralities. The action is just the sum of two Weyl fermion actions. It
seems, however, non-trivial to get the action of the WZW model from two
chiral boson actions of opposite ``chiralities'', because it is not the
direct sum of two chiral bosons.

To solve this problem, Stone \cite{ms} introduced the idea of soldering the
two chiral scalars by introducing a non-dynamical gauge field to remove the
degree of freedom that obstructs the vector gauge invariance \cite{w2}. This
is connected, as we said above, to the necessity that one must have more
than the direct sum of two fermions representations of the Kac-Moody algebra
to describe a Dirac fermion. In another way we can say that the equality for
the weights in the two representations is physically connected with the
necessity to abandon one of the two separate chiral symmetries, and accept
that a non-chiral gauge symmetry should be kept. This is the main motivation
for the introduction of the soldering field which makes possible the fusion
of dualities in all space-time dimensions. This restriction will force the
two independent chiral representations to belong to the same multiplet,
effectively soldering them together.

The basic idea of the soldering procedure is to raise a global
Noether symmetry of the self and antiself dual constituents into a
local one. The effective theory, consists of the dual components
and an interference term \cite{w2}.

An iterative Noether procedure \cite{w2} is usually adopted in order to
promote global symmetries. Therefore, one supposes that the symmetries in
question are being described by the local actions, invariant under a global
transformation. Then, trying to raise the symmetry to a local one, notice
that now under local transformations these actions will not remain
invariant, and Noether counter-terms become necessary to reestablish the
invariance, along with appropriate auxiliary fields, the so-called soldering
fields which by construction should be non-dynamical ones.

For each the self and antiself dual system we have in mind that
this iterative gauging procedure is constructed not to produce
invariant actions for any finite number of steps. However, if
after N repetitions, the non invariant piece ends up being only
dependent on the gauging parameters and Noether currents, then
there will exist the possibility of mutual cancelation if both
self and antiself gauged systems are combined with each other.

Finally, the auxiliary fields should be eliminated, for instance,
through its equations of motion, from the resulting effective
action, in favor of the physically relevant degrees of freedom. It
is important to notice that after the elimination of the soldering
fields, the resulting effective action will not depend on either
self or antiself dual fields but only on some collective field,
defined in terms of the original ones in a invariant way.

\section{The chiral Schwinger model}

Let us begin by introducing the notation used here for the light cone
variables:
\begin{eqnarray}
x_{\pm } &=&{\frac{1}{\sqrt{2}}}(x_{0}\pm x_{1})\;\;,  \nonumber \\
\partial _{\pm } &=&{\frac{1}{\sqrt{2}}}(\partial _{0}\,\pm \,\partial
_{1})\;\;,  \nonumber \\
A_{\pm } &=&{\frac{1}{\sqrt{2}}}(A_{0}\,\pm \,A_{1})\;\;,
\end{eqnarray}
and now we can work out our model.

We can write down the interaction terms of the chiral Schwinger model for
both chiralities in its bosonized form as:
\begin{equation}
W_{+}\,=\,\int \,d^{2}\,x\left( \partial _{+}\phi \,\partial _{-}\phi
\,+\,\,e\,A_{+}\,\partial _{-}\phi \,+\,\frac{a\,e^{2}}{4}%
\,A_{+}\,A_{-}\right) \label{wp}\;\;,
\end{equation}
and
\begin{equation}
W_{-}\,=\,\,\int \,d^{2}\,x\left( \partial _{+}\rho \,\partial _{-}\rho
\,+\,\,e\,A_{-}\,\partial _{+}\rho \,+\,\frac{b\,e^{2}}{4}%
\,A_{+}\,A_{-}\right) \label{wm}\;\;,
\end{equation}
where $a$ and $b$ are the Jackiw-Rajaraman coefficients for each chirality
respectively \cite{jr}. Notice that $W_{+}$ and $W_{-}$ are invariant under
the following semi-local gauge transformations respectively

\begin{equation}
\delta A_{+}=0\,;\,\delta A_{-}=\partial _{-}\epsilon _{R}\,;\,\delta \phi
=-a\,\,e\,\,\epsilon _{R}/4  \label{semi1}
\end{equation}

\begin{equation}
\delta A_{+}=\partial _{+}\epsilon _{L}\,;\,\delta A_{-}=0\,;\,\delta \phi
=-b\,e\,\epsilon _{L}/4  \label{semi2}
\end{equation}

\noindent where $\epsilon_R = \epsilon_R (x_-)$ and $\epsilon_L = \epsilon_L
(x_+)$. Performing a direct sum of the actions we have:

\begin{eqnarray}
W_{TOTAL}\, &=&\,W_{+}\,\oplus \,W_{-}\,  \nonumber  \label{total} \\
&=&\,\,\int \,d^{2}\,x\,\left( \,\partial _{+}\phi \,\partial _{-}\phi
\,+\,\partial _{+}\rho \,\partial _{-}\rho \,+\,2\,e\,A_{+}\,\partial
_{-}\phi \,+\,2\,e\,A_{-}\,\partial _{+}\rho \right.  \nonumber \\
&&\left. \,+\,\frac{(a+b)\,e^{2}}{4}\,A_{+}\,A_{-}\,\right) \;\;,
\end{eqnarray}
Notice that if $a+b=2$, $W_{TOTAL}$ is invariant under $\delta \rho
=\epsilon =\delta \sigma ;\,\delta A_{\pm }=-2\,\partial _{\pm }\epsilon $
where $\epsilon =\epsilon (x_{\mu })$ is an arbitrary function. In order to
make this local gauge invariance explicit, let us do the following rotation

\begin{eqnarray}
\sqrt{2}\,\rho &=&\sigma \,+\,\varphi \\
\sqrt{2}\,\phi &=&\sigma \,-\,\varphi \;\;.
\end{eqnarray}
Substituting in (\ref{total}) and writing the result in a explicit covariant
way we have a vector plus an axial Schwinger model:
\begin{equation}
\mathcal{L}\,=\mathcal{L}_{VSM}\,+\mathcal{\,L}_{ASM}\,+\,{\frac{1}{4}}%
\,(\,a\,+\,b\,-2)\,e^{2}\,A_{\mu }^{2}  \label{directsum}
\end{equation}
with
\begin{eqnarray}
\mathcal{L}_{VSM}\left( \sigma ,A_{\mu }\right) &=&{\frac{1}{2}}\,(\partial
_{\mu }\,\sigma )^{2}\,+\,e\,\varepsilon ^{\mu \nu }\,\partial _{\mu
}\,\sigma \,A_{\nu }\,,  \nonumber \\
&& \\
\mathcal{L}_{ASM}\left( \phi ,A_{\mu }\right) &=&{\frac{1}{2}}\,(\partial
_{\mu }\,\phi )^{2}\,+\,e\,g^{\mu \nu }\,\partial _{\mu }\,\sigma \,A_{\nu
}\,+\frac{1}{2}A_{\mu }^{2},  \nonumber
\end{eqnarray}
The gauge invariance for $a+b=2$ is now explicit.

At this point we observe that, contrary to what one might think
based on claims in \cite{w2}, \cite{bw527}, it is not really
necessary to perform the soldering of the two chiral bosons in
order to accomplish the full local gauge invariance. Since the
usual soldering procedure obtained after the addition of an
interference term, produces only the vector Schwinger model, the
result (\ref{directsum}) suggests the existence of a second
soldering procedure giving rise to the axial Schwinger model.
Indeed, that is what we have found as we next show.

The actions $W_{+}$ and $W_{-}$ only depend on $\phi $ and $\rho $ through
derivatives and therefore are invariant under rigid translations of these
fields, the basic idea in the soldering procedure is to join the actions $%
W_{+}$ and $W_{-}$ into a new one while promoting this symmetry to a local
form. Let us suppose the local variations
\begin{equation}
\delta \phi =\eta \left( x\right) ;\,\delta \rho =\alpha \,\eta \left(
x\right) ,  \label{1}
\end{equation}

\noindent where $\alpha $ is, at this point, an arbitrary constant. In the
usual soldering procedure one assumes $\alpha =1$. Under (\ref{1}) we have:
\begin{equation}
\delta \left( W_{+}+W_{-}\right) =\int \left( J_{+}\,\partial _{-}\eta
+J_{-}\,\partial _{+}\eta \right) d^{3}x,
\end{equation}

\noindent with
\begin{equation}
J_{+}=\partial _{+}\phi \,+\,e\,A_{+},\,\,J_{-}=\alpha \,\left( \partial
_{-}\rho \,+\,e\,A_{-}\right) .
\end{equation}

\noindent Introducing two auxiliary fields $B_{\pm }$ such that
\begin{equation}
\delta B_{\pm }=-\,\partial _{\pm }\eta ,  \label{4}
\end{equation}
we have
\begin{eqnarray}
\delta \left( \mathcal{L}_{+}+\mathcal{L}_{-}+B_{+}\,J_{-}+B_{-}\,J_{+}%
\right) &=&B_{-}\partial _{+}\eta + \alpha^2 B_{+}\partial _{-}\eta =
\nonumber \\
&& \\
&=&-\,\delta \left( B_{+}B_{-}\right) +\left( 1-\alpha ^{2}\right)
\,B_{+}\,\delta B_{-}\,.  \nonumber
\end{eqnarray}

\noindent Therefore, for $\alpha =\pm 1$ we can define a soldered Lagrangian
density, invariant under (\ref{1}) and (\ref{4}), which is given by
\begin{equation}
\mathcal{L}_{\alpha }^{\left( s\right) }=\mathcal{L}_{+}+\mathcal{L}%
_{-}+B_{+}\,J_{-}+B_{-}\,J_{+}+B_{+}\,B_{-}.
\end{equation}

\noindent Eliminating the auxiliary fields through their equations of
motion, we have
\begin{eqnarray}
\mathcal{L}_{\alpha }^{\left( s\right) } &=&\mathcal{L}_{+}+\mathcal{L}%
_{-}-J_{+}\,J_{-}=  \nonumber \\
&&  \label{7} \\
&=&\partial _{+}\Phi \,\partial _{-}\Phi +e\,\left( A_{+}\partial _{-}\Phi
-\alpha \,A_{-}\partial _{+}\Phi \right) +\frac{e^{2}\left( a+b-2\alpha
\right) }{4}A_{+}A_{-},  \nonumber
\end{eqnarray}

\noindent where $\Phi =\phi -\alpha \,\rho $ is a field combination
invariant under (\ref{1}). In both cases $\alpha =+1$ and $\alpha =-1$, if
we choose $a+b=2$, we recover the vector and the axial Schwinger models
respectively.
\begin{eqnarray}
\mathcal{L}_{+1}^{\left( s\right) } &=&{\frac{1}{2}}\,(\partial _{\mu
}\,\Phi )^{2}\,+\,e\,\varepsilon ^{\mu \nu }\,\partial _{\mu }\,\Phi
\,A_{\nu }\,,  \nonumber \\
&&  \label{dual} \\
\mathcal{L}_{-1}^{\left( s\right) } &=&\frac{1}{2}\left( \partial ^{\mu
}\Phi +e\,A^{\mu }\right) ^{2}.  \nonumber
\end{eqnarray}

Thus, the new soldering found here for $\alpha =-1$ generates the missing
part in the rotated direct sum (\ref{directsum}). In the two cases $\alpha
=\pm 1$, the soldering procedure has produced a Lorentz covariant and local
gauge invariant Lagrangian out of two anomalous gauge models which possessed
only a semi-local gauge invariance. It is remarkable that in order to prove
gauge invariance of $\mathcal{L}_{-1}^{\left( s\right) }$ under $\delta
A_{\mu }=\partial _{\mu }\eta $ we also need to vary the scalar composite
field $\delta \Phi =-e\,\eta $ which was invariant from the start under the
transformation (\ref{1}). Even in the case of $\mathcal{L}_{+1}^{\left(
s\right) }$ it is assumed in the soldering approach \cite{abw} that $\delta
A_{\mu }=0$ and one ends up with a local gauge invariance with $\delta
A_{\mu }=\partial _{\mu }\eta \neq 0$. Thus, the connection between the
original soldering symmetry and the final gauge symmetry is rather
mysterious in both cases $\alpha =\pm 1$. We will see later that the same
phenomenon appears in $d=3$ dimensions. Regarding the choice $a+b=2$ we
point out that it is in agreement with the unitarity bound $a\geq 1$ and $%
b\geq 1$ of the chiral Schwinger models. As argued in \cite{abw}, the same result
$a=1=b$ could be found by using a left/right symmetry ($\pm \leftrightarrow \mp $),
implying that $a=b$ in the chiral Lagrangian densities and the condition $a+b=2$, with
no need of unitarity arguments. We remark that although the models (\ref{dual}) are
not equivalent, by
functionally integrating the soldering field $\Phi $ in (\ref{7}) and using $%
a+b=2$ we obtain a $\alpha $-independent gauge invariant effective action
which, after adding the Maxwell Lagrangian density, describes a massive
photon:
\begin{equation}
\mathcal{L}_{eff}\left[ A_{\mu }\right] =-\frac{1}{4}F^{\mu \nu }\frac{%
\left( \Box +e^{2}/\pi \right) }{\Box }F_{\mu \nu }.
\end{equation}

\noindent It is important to point out that one of the motivations to introduce the
usual soldering procedure ($\alpha = 1$) is to offer an explanation, see \cite{abw},
for the formal chiral factorization identity of fermionic determinants:

\be \det \left(i\dslash + e \aaslash \right) = \det \left(i\dslash + e \aaslash_+
\right)  \det \left(i\dslash + e \aaslash_- \right) \label{id1} \ee

\no Where $\aaslash_{\pm} = \aaslash P_{\pm} $ with $ P_{\pm} = (1 \pm \gamma_5)/2$. A
trivial direct sum of the two bosonized versions of the chiral determinants gives rise
to a sum of the axial and vector Schwinger model and therefore does not reproduce the
bosonized version of the vector Schwinger model only, while this is achieved by the
usual soldering procedure. On the other hand, by tracing back the second soldering
procedure ( $\alpha=-1$ ) it is easy to show that it is technically equivalent to the
usual soldering of $W_+$ and $W_-(-e)$ where $W_-(-e)$ corresponds to change the sign
of the charge in $W_-$. Since we ended up with the axial Schwinger model, there should
be the factorization formula:

\be \det \left(i\dslash + e \aaslash \gamma_5 \right) = \det \left(i\dslash + e
\aaslash_+ \right)  \det \left(i\dslash - e \aaslash_- \right) \label{id2} \ee

\no Indeed, by splitting the fermions in chiral components it is easy to derive the
above identity. Therefore the second soldering has helped us in finding factorization
formulas for the fermionic determinant which may be useful in other applications as in
$D=3$ where bosonization is much less developed.

Concluding, the generalized soldering procedure has produced in $%
D=2$ a self-consitent Lorentz covariant theory with local gauge invariance for
$\alpha=\pm 1$ in agreement with our expectations based on the direct sum of the two
chiral Schwinger models.

\section{D=3}

Now let us explore the generalized soldering in the higher
dimensional case of two Maxwell-Chern-Simons models with opposite
sign masses in the presence of a nonminimal iteraction:

\begin{equation}
W_+ = \int d^3x\left\lbrack -\frac 14 F_{\mu\nu}^2(A) + \frac{m_+}2
\epsilon_{\alpha\beta\gamma}A^{\alpha}\partial^{\beta}A^{\gamma} + \gamma_+
\epsilon_{\alpha\beta\gamma}J^{\alpha}\partial^{\beta}A^{\gamma}
\right\rbrack  \label{wmais3}
\end{equation}

\begin{equation}
W_- = \int d^3x\left\lbrack -\frac 14 F_{\mu\nu}^2(B) - \frac{m_-}2
\epsilon_{\alpha\beta\gamma}B^{\alpha}\partial^{\beta}B^{\gamma} + \gamma_-
\epsilon_{\alpha\beta\gamma}J^{\alpha}\partial^{\beta}B^{\gamma}
\right\rbrack  \label{wmenos3}
\end{equation}

\noindent The above theories have been considered before \cite{bk}, with $%
J_{\nu }=0$, in its dual form which corresponds to a self-dual and
an antiself-dual model. In \cite{bk} the models (\ref{wmais3}) and
(\ref {wmenos3})have been soldered in the usual way ($\alpha =1$)
into a Maxwell-Chern-Simons-Proca theory. The general case
$m_{+}\ne m_{-}$ required the use of the equations of motion
during the soldering procedure while in the special case
$m_{+}=m_{-}$ the whole method works off-shell. Such special case
naturally appears in the bosonization of $QED_{3}$ with two
components fermions in the large mass limit. It has been first
considered from the soldering point of view in \cite{bw527} where
it was shown that the parity non-invariant theories
(\ref{wmais3}), (\ref{wmenos3}) are soldered into a parity
invariant Maxwell theory with a Proca term. The soldered theory is
a function of the composite field $A_{\mu }-B_{\mu }$. The
nonminimal couplings $\gamma _{\pm }$ introduced here allow a
generalization of \cite{bk,bw527} to an interacting theory. In this sense $%
J_{\mu }$ play a role similar to the gauge field in the chiral
Schwinger models of last section. We have chosen a nonminimal
coupling with the current $J_{\mu }$ because it keeps the theory
invariant under rigid translations of the gauge field which will
play an important role in the soldering mechanism. Besides, the
nonminimal coupling naturally appears when we search for the dual
of the self-dual model minimally coupled to a current as obtained
in \cite{GMS}, see also \cite{Anacleto2,jpa}. In fact in that case
we have $\gamma _{\pm }=\pm e/m_{\pm }$ where $e$ is the coupling
appearing in the minimal coupling term. Now we start our
generalized soldering procedure by lifting the rigid translation
symmetry of (\ref {wmais3})and (\ref{wmenos3}) to the local form:

\begin{equation}
\delta A_{\mu} = \eta_{\mu} \quad ; \quad \delta B_{\mu} = \alpha \eta_{\mu}
\label{13}
\end{equation}

\noindent which imply

\begin{equation}
\delta \left( W_+ + W_- \right) = \int d^3 x J_{\mu\nu}
\partial^{\mu}\eta^{\nu}  \label{14}
\end{equation}

\noindent where

\begin{equation}
J_{\mu\nu} = - F_{\mu\nu} (A) - \alpha F_{\mu\nu} (B) +
\epsilon_{\mu\nu\gamma}\left(m_+ A^{\gamma} - \alpha m_- B^{\gamma}\right)
 + (\alpha\gamma_- + \gamma_+)\epsilon_{\mu\nu\gamma}J^{\gamma}
\label{15}
\end{equation}

\noindent As usually one introduces auxiliary fields $B_{\mu\nu}$ such that

\begin{equation}
\delta B_{\mu\nu} = - \partial_{\mu} \eta_{\nu}  \label{15b}
\end{equation}

\noindent It is easy to derive:

\begin{equation}
\delta \left(W_+ + W_- + \int d^3x B_{\mu\nu}J^{\mu\nu} \right) = \delta
\left\lbrack \frac{(1+\alpha^2)}2 \int d^3x B_{\mu\nu}(B^{\mu\nu} -
B^{\nu\mu}) \right\rbrack + \int d^3x \epsilon_{\mu\nu\gamma}(m_+-\alpha^2
m_-)\eta^{\gamma}  \label{16}
\end{equation}

\noindent Since $\eta^{\gamma}$ can not be written as a local function of $%
\delta B_{\mu\nu}$ we choose at this point

\begin{equation}
\alpha = \pm \sqrt{\frac {m_+}{m_-}}  \label{17}
\end{equation}

\noindent Consequently we are able to build an invariant action under the
local translations (\ref{13}). After the elimination of the auxiliary fields
$B_{\mu\nu}$ through their equations of motion the reader can check that we
arrive at:

\begin{equation}
W_{\alpha}^{(S)} = W_+ + W_- + \int d^3x \frac{J^{\mu\nu}J_{\mu\nu}}{%
4(1+\alpha^2)} \equiv \int d^3x \mathcal{L}_{\alpha}^{(S)}  \label{18}
\end{equation}

\noindent where after some rearrangements we can write down:

\bea
\mathcal{L}_{\alpha }^{(S)} = &-&\frac{F_{\mu \nu }^{2}(C)}{4(1+\alpha ^{2})}+%
\frac{(m_{+}-m_-)}{2(1+\alpha ^{2})}\epsilon _{\mu \nu \gamma
}C^{\mu
}\partial ^{\nu }C^{\gamma }+\frac{1}{2(1+ \alpha^2)}\left\lbrack \alpha m_{-}C_{\mu }+%
(\alpha\gamma _{-} +  \gamma _{+})\,J_{\mu }\right\rbrack^{2}\nn\\
&+& \frac{(\alpha\gamma_+ - \gamma_-)}{1+\alpha^2}\epsilon _{\mu
\nu \gamma}J^{\mu}\partial^{\nu}C^{\gamma}\label{19} \eea

\noindent The composite soldering field $C_{\mu} = \alpha A_{\mu} - B_{\mu}$
is invariant under the local transformations (\ref{13}). Observe that no use
has been made of the equations of motion contrary to \cite{bk}. In (\ref{19}%
) we still have two choices $\alpha=\pm \sqrt{m_+/m_-}$.
Suppressing the interaction $J_{\mu}=0$ we recover the same
Maxwell-Chern-Simons-Proca of \cite{bk}. In parallel with the
$d=2$ results of last section , the dependence on $\alpha$ only
appears through the interacting terms which now contain a minimal
coupling $J\cdot C$, a Thirring term $J^2$ and a nonminimal
coupling of the Pauli type $\epsilon _{\mu \nu
\gamma}J^{\mu}\partial^{\nu}C^{\gamma}$. Besides the soldering
symmetry (\ref{13}), if we think of $J_{\mu}$ as a dynamical
field, the soldered Lagrangian (\ref{19}) is symmetric under:

\begin{eqnarray}
J_{\mu} &\to & J_{\mu} - \partial_{\mu} \eta  \nonumber \\
C_{\mu} &\to & C_{\mu} + \frac{\alpha}{m_+} (\alpha \gamma_- +
\gamma_+)\partial_{\mu}\eta  \label{20}
\end{eqnarray}

\noindent For $\alpha\gamma_- + \gamma_+ \ne 0 $ the symmetry (\ref{20}) is
larger than the originally envisaged soldering symmetry (\ref{13}) in
analogy with the case $\alpha=-1$ of last section. In the $d=2$ example the
two soldering choices $\alpha=\pm 1$ have led to different actions but the
same effective action for the electromagnetic field $A_{\mu}$. Now in $d=3$
if we Gaussian integrate over the soldering field $C_{\mu}$ in the path
integral we derive from (\ref{19}) the effective Lagrangian:

\begin{eqnarray}
\mathcal{L}_{eff}^{(S)}[J] &=&\frac{m_{-}\left( \gamma _{+}^{2}+\gamma
_{-}^{2}\right) J^{\mu }(\Box +m_{+}m_{-})\Box \theta _{\mu \nu }J^{\nu }}{%
(m_{+}+m_{-})\left[ \left( \Box +m_{+}m_{-}\right)
^{2}+(m_{+}-m_{-})^{2}\Box \right] }+  \nonumber \\
&&+\frac{m_{-}(m_{+}-m_{-})(\alpha \gamma _{-}+\gamma _{+})^{2}J^{2}}{%
2(m_{+}+m_{-})^{2}}\,+  \label{21} \\
&+&\frac{m_{-}J^{\mu }\left[ (m_{-}\gamma _{-}^{2}-m_{+}\gamma _{+}^{2})\Box
+m_{+}m_{-}(m_{+}\gamma _{-}^{2}-m_{-}\gamma _{+}^{2})\right] \epsilon _{\mu
\nu \gamma }\partial ^{\nu }J^{\gamma }}{(m_{+}+m_{-})\left[ \left( \Box
+m_{+}m_{-}\right) ^{2}+(m_{+}-m_{-})^{2}\Box \right] }  \nonumber
\end{eqnarray}

\noindent where $\theta_{\mu\nu}=g_{\mu\nu}-\partial_{\mu}\partial_{\nu}/%
\Box $. Thus, in general the two possible choices for $\alpha$ lead us to
inequivalent theories due to the second term in (\ref{21}) even after
integration over the soldering field. Only in the special case $%
m_+=m_-\equiv m (\alpha^2=1)$ we have the same result for both choices $%
\alpha = \pm 1$, although the Lagrangians before the integration over the
soldering field are different for $\alpha=1$ and $\alpha=-1$. In particular,
for $\gamma_-=-\gamma_+ \equiv \gamma$ we have

\begin{equation}
\mathcal{L}_{\alpha=1}^{(S)} = - \frac{F_{\mu\nu}^2}8 + \frac{m^2 C^2}4 -
\gamma \epsilon_{\mu\nu\gamma}J^{\mu}\partial^{\nu}C^{\gamma}  \label{21b}
\end{equation}

\begin{equation}
\mathcal{L}_{\alpha=-1}^{(S)} = - \frac{F_{\mu\nu}^2}8 + \frac{m^2 C^2}4 +
m\gamma J\cdot C + \gamma^2 J^2  \label{21c}
\end{equation}

\noindent We see that a nonminimal coupling of the Pauli type for
$\alpha=1$ is traded in a nonminimal coupling plus a Thirring term
for $\alpha=-1$. It deserves a comment the fact that such
correspondence is known to appear in the dual theory of a
self-dual model minimally coupled to $U(1)$ matter fields, see
\cite{GMS,Anacleto2,jpa}, which corresponds to a
Maxwell-Chern-Simons theory nonminimally coupled to $U(1)$ matter
fields plus a Thirring interaction. The similarity with the $d=2$
case is remarkable. In that case the interaction term $e
\epsilon^{\mu\nu}\partial_{\mu}\Phi A_{\nu}$ has been traded in $%
e\partial_{\mu}\Phi A^{\mu} + e^2 A^2/2$ as if the complementary solderings $%
\alpha=1$ and $\alpha=-1$ were generating dual theories. This point
certainly demands a deeper investigation.

It is interesting to note that when we drop the spatial dependence of the
vector fields in (\ref{wmais3}) and (\ref{wmenos3}) and set $\gamma _{\pm
}=0 $ we recover the Lagrangians $\mathcal{L}_{\pm }=(1/2)\left[ \dot{x}%
_{i\pm }\pm \omega _{\pm }\epsilon _{ij}x_{i\pm }\dot{x}_{j\pm }\right] $
where we have relabelled $A_{i}\to x_{i+}\,;\,B_{i}\to x_{i-}$ and $m_{\pm
}\to \omega _{\pm }$. The Lagrangians $\mathcal{L}_{+}$ and $\mathcal{L}_{-}$
describe a right and left moving particle on a plane in the presence of a
constant magnetic field orthogonal to the plane and pointing in opposite
directions respectively. Those Lagrangians have been considered before in
\cite{djt,bw527,bg} and soldered in \cite{bk} where use has been made of the
equations of motion for $\omega _{+}\ne \omega _{-}$. If we choose $\delta
x_{i+}=\eta _{i}$ and $\delta x_{i-}=\alpha \eta _{i}$ with $\alpha
^{2}=\omega _{+}/\omega _{-}$ we end up with a soldered Lagrangian which
represents a two dimensional Harmonic oscillator in the presence of a
residual magnetic field which disappears for $m_{+}=m_{-}$, i.e., in terms
of the soldering field $\Phi _{i}=\left( \alpha x_{i+}-x_{i-}\right) /\sqrt{%
1+\alpha ^{2}}$ one has $\mathcal{L}^{(S)}=(1/2)\left[ \dot{\Phi}%
^{2}+(\omega _{+}-\omega _{-})\epsilon ^{ij}\Phi _{i}\dot{\Phi}_{j}-\omega
_{+}\omega _{-}\Phi _{i}^{2}\right] $ in agreement with the final result of
\cite{bk} but we stress that our generalized soldering differently from \cite
{bk} does not require the use of equations of motion, except of course for
the auxiliary fields whose equations of motion are actually mathematical
identities without dynamical content.

\section{Conclusion}

Chiral Schwinger models possess only a semilocal form of gauge
invariance. However, it is known that one can recover full local
gauge invariance by soldering two opposite chirality chiral
Schwinger models. In this case one ends up with a vector Schwinger
model. We have shown here that a direct sum of both chiral models
already furnishes a local gauge invariant theory with no need of
soldering. After a simple rotation of the fields the resulting
theory becomes a vector plus an axial Schwinger model so
suggesting the existence of a second soldering procedure leading
us to the vector Schwinger model. Indeed, we have found here a
generalization of the soldering mechanism leading either to the
axial or to the vector models depending on a constant parameter
$\alpha = \pm 1$. The same twofold generalization occurs in $d=3$
when we solder two Maxwell-Chern-Simons theories with opposite
sign masses $m_+$ and $-m_-$. In this case we have $\alpha = \pm
\sqrt{m_+/m_-}$. By allowing $\alpha\ne 1$ we have been able to
implement the soldering algorithm off-shell even for $m_+\ne m_-$
so generalizing previous results in the literature. Introducing a
nonminimal interaction of the Maxwell-Chern-Simons fields with a
vector current we obtain different soldered theories for the two
different choices for $\alpha$. However, in the special case
$\alpha=\pm 1$ in both $d=2$ and $d=3$ we have the same
theory after integration over the soldering field. Apparently, the cases $%
\alpha=1$ and $\alpha=-1$ generate dual versions of the same theory as we
have argued in the last section. For the new soldering found here ($%
\alpha=-1 $) the resulting gauge symmetry is surprisingly larger than the
originally imposed soldering symmetry. In order to get a deeper
understanding of this and other aspects of such complementary solderings we
are now applying it to the cases of non-abelian gauge theories in $d=2$,
models including coupling to 2d gravity and electromagnetic duality in $d=4$
which have all been considered from the point of view of usual soldering $%
(\alpha=1)$. Clearly, a better understanding of the generalized soldering
from the path integral point of view through a possible master action
approach would probably unravel some of the interesting features of this
mechanism.

\newpage

\section{Acknowledgments}

This work is partially supported by Conselho Nacional de Pesquisa e
Desenvolvimento (CNPq).


\begin{thebibliography}{99}
\bibitem{gross}  D. J. Gross, J. A. Harvey, E. Martinec and R. Rohm, Phys.
Rev. Lett. \textbf{54} (1985) 502; Nucl. Phys. B \textbf{256} (1985) 253;
\textbf{267} (1986) 75.

\bibitem{grs}  M. Gomes, V. Kurak, V. O. Rivelles and A. J. da Silva Phys.
Rev. D \textbf{38} (1988) 1344.

\bibitem{siegel}  W. Siegel, Nucl. Phys. B \textbf{238} (1984) 307.

\bibitem{fj}  R. Floreanini and R. Jackiw, Phys. Rev. Lett. \textbf{59}
(1987) 1873.

\bibitem{ws}  X. G. Wen, Phys. Rev. Lett. 64 (1990) 2206; Phys. Rev. B
\textbf{41} (1990) 12838; M. Stone, Phys. Rev B \textbf{42} (1990)
212.

\bibitem{ggkrs}  H. O. Girotti, M. Gomes, V. Kurak, V. O. Rivelles and A. J.
da Silva, Phys. Rev. Lett. \textbf{60} (1988) 1913.


\bibitem{ms}  M. Stone, Illinois preprint, ILL - (TH) - 89-23, 1989; Phys.
Rev. Lett. \textbf{63} (1989) 731; Nucl. Phys. B \textbf{327}
(1989) 399; D. Depireux, S. J. Gates Jr. and Q-Han Park, Phys.
Lett. B \textbf{224} (1989) 364; E. Witten, Commun. Math. Phys.
\textbf{144} (1992) 189.

\bibitem{orbitas}  A. A. Kirilov, \textit{Elements of the representation
theory}, Nakua, Moscow, 1972; A. Alekseev, L. Faddeev and S.
Shatashvili, J. Geom. and Phys. \textbf{5 }(1988) 391; A. Alekseev
and S. Shatashvili, Nucl. Phys. B \textbf{323} (1989) 719.

\bibitem{wzw}  E.Witten, Comm. Math. Phys. \textbf{92} (1984) 455.

\bibitem{adw}  R. Amorim, A. Das and C. Wotzasek, Phys. Rev. D \textbf{53}
(1996) 5810.

\bibitem{abw}  E. M. C. Abreu, R. Banerjee and C. Wotzasek, Nucl. Phys. B
\textbf{509} (1998) 519.

\bibitem{schw}  J. Schwinger, Phys. Rev. \textbf{128} (1962) 2425.

\bibitem{jr}  R. Jackiw and R. Rajaraman, Phys. Rev. Lett. \textbf{54}
(1985) 1219.

\bibitem{aw}  E. M. C. Abreu and C. Wotzasek, Phys. Rev. D \textbf{58}
(1998) 101701.

\bibitem{gs}  S. J. Gates, Jr. and W. Siegel, Phys. Lett. B \textbf{206}
(1988) 631.

\bibitem{hull}  C. M. Hull, Phys. Lett. B \textbf{206} (1988) 234; Phys.
Lett. B \textbf{212 (}1988) 437.

\bibitem{tpn}  P. K. Townsend, K. Pilch and P. van Nieuwvenhuizen, Phys.
Lett. B \textbf{136} (1984) 38.

\bibitem{dj}  S. Deser and R. Jackiw, Phys. Lett. B \textbf{139} (1984) 371.

\bibitem{bw527}  R. Banerjee and C. Wotzasek, Nucl. Phys. B \textbf{527}
(1998) 402

\bibitem{ss}  J. Schwarz and A. Sen, Nucl. Phys. B \textbf{411} (1994) 35.

\bibitem{dutra}  A. de Souza Dutra, Phys. Lett. B \textbf{286} (1992) 285.

\bibitem{ps}  A. Pressley and G. Segal, ``Loup Groups'', Oxford University
Press, Oxford, 1986.

\bibitem{w2}  For a review see: C. Wotzasek, ``Soldering Formalism: theory
and applications'', hep-th 9806005.

\bibitem{bk}  R. Banerjee and S. Kumar, Phys. Rev. D \textbf{60}, (1999)
085005.

\bibitem{GMS}  M. Gomes, L. C. Malacarne and A. J. da Silva, Phys. Lett. B
\textbf{439} (1998) 137.

\bibitem{Anacleto2}  M.A. Anacleto, A. Ilha, J.R.S. Nascimento, R.F. Ribeiro
and C. Wotzasek, Phys. Lett. B \textbf{504} (2001) 268.

\bibitem{jpa}  D. Dalmazi, Journal of Phys. A \textbf{37} (2004) 2487.

\bibitem{djt}  G.V. Dunne, R. Jackiw and C.A. Trugenberger, Phys. Rev. D
\textbf{41} (1990) 661.

\bibitem{bg}  R. Banerjee and S. Ghosh, J. of Phys. A \textbf{31} (1998)
L603.

\end{thebibliography}
\end{document}